# A GENERAL RELATIVISTIC MODEL FOR MAGNETIC MONOPOLE-INFUSED COMPACT OBJECTS

Z. Pazameta
Physical Sciences Department
Eastern Connecticut State University
Willimantic CT 06226
USA

pazameta@easternct.edu

ABSTRACT

Emergent concepts from astroparticle physics are incorporated into a classical solution of the Einstein-Maxwell equations for a binary magnetohydrodynamic fluid, in order to describe the final equilibrium state of compact objects infused with magnetic monopoles produced by proton-proton collisions within the intense dipolar magnetic fields generated by these objects during their collapse. It is found that the effective mass of such an object's acquired monopolar magnetic field is three times greater than the mass of its native fluid and monopoles combined, necessitating that the interior matter undergo a transition to a state of negative pressure in order to attain equilibrium. Assuming full symmetry between the electric and magnetic Maxwell equations yields expressions for the monopole charge density and magnetic field by direct analogy with their electrostatic equivalents; inserting these into the Einstein equations then leads to an interior metric which is well-behaved from the origin to the surface, where it matches smoothly to an exterior magnetic Reissner-Nordström metric free of any coordinate pathologies. The source fields comprising the model are all described by simple, well-behaved polynomial functions of the radial coordinate, and are combined with straightforward regularity conditions to yield expressions delimiting several fundamental physical parameters pertaining to this hypothetical astrophysical object.

## 1. INTRODUCTION

During the past decade or so there has been a resurgence of activity in the theoretical study of astrophysical compact objects, resulting in a variety of models for such entities in what we may call the "compact-object desert" between the neutron-star state and putative total collapse to a singularity. One reason for this is increasing interest in the astrophysical implications of exotic entities such as dark matter, dark energy, and vacuum energy. Another stems from novel applications of fundamental concepts of quantum theory to micromatter, which has inspired investigators to speculate that even quarks may not be elementary but may themselves be composed of sub-particles called preons—and these preons may in turn be composed of smaller entities, and so on. This hierarchy would be expected to terminate at the smallest physically meaningful length-scale in quantum theory, the Planck length $l_P \equiv \sqrt{(\hbar G/c^3)} \approx 10^{-35}$ m, which is considered to be the lower bound for the size of any physical entity (whether elementary or composite); one may therefore postulate the existence of a variety of new compact objects (Hansson, 2007).

The model in this paper describes a class of such hypothetical compact objects, those which have produced and accumulated in their interiors a population of magnetic monopoles; for convenience, we give such an object the name, "monostar."

Consider a collapsing stellar remnant, with a mass similar to or greater than the upper limit for the mass of a neutron star. As it collapses, the intense dipolar magnetic field generated by the remnant's rapid rotation and internal currents leads to the production of magnetic monopole-antimonopole pairs via photon-photon fusion in ultra-high energy proton-proton collisions (Dougall & Wick, 2008). As has been proposed for pulsars (Bonnardeau & Drukier, 1979), the magnetic field pulls one component of this pair into the collapsing object's interior and ejects the other into interstellar space. Since energy is removed from the magnetic field by this process, it is expected that both the field and the object's rotation will decay as the collapse proceeds.

Our model represents the terminal state of this collapse scenario, where the compact object ends up with zero rotation and no self-generated (dipolar) magnetic field—but has acquired a monopolar magnetic field from its accumulated monopoles, with this field playing a major part in defining the object's final equilibrium state. The model is based on a binary fluid composed of two non-interacting fermion particle species; one component represents the monostar's native matter as a simple perfect fluid, and the other (magnetohydrodynamic) component represents the infused monopoles.

## 2. THE GEOMETRY

The monostar is considered to be spherically symmetric, non-rotating (stationary) and fully evolved (static), so both the interior and exterior geometries may be described by static, spherically symmetric metrics that are functions of the radial coordinate only.

By Birkhoff's theorem, the exterior metric for such an object must have the form

$$ds^2 = f.dt^2 - (1/f).dr^2 - r^2 d\Omega^2, \qquad (1)$$

where $f = f(r)$ only and $d\Omega^2 \equiv d\theta^2 + \sin^2\theta.d\varphi^2$. (We set $\theta = \pi/2$ and $c = G = 1$ throughout.) We shall assume that the interior metric also has the form of relation (1), for three reasons: It will be easier to match to the exterior metric across the boundary; the Einstein equations become considerably simpler; and, above all, it turns out that the physics requires it: Tiwari, Kanakamedala & Rao (1984) showed that an interior static, spherically symmetric metric must be of the form given by expression (1) in order that a sphere of electrically charged fluid particles shall be in equilibrium, and we argue below that this result also holds if magnetic charges are substituted for the electric ones.

The obvious choice to describe the monostar's exterior gravitational field is the Reissner-Nordström metric

$$ds^2 = \Phi.dt^2 - (1/\Phi).dr^2 - r^2 d\Omega^2, \qquad (2)$$

where $\Phi \equiv 1 - r_S/r + Q_m^2/r^2$ with $Q_m$ being the object's global magnetic charge and $r_S$ its Schwarzschild radius (which, at this point, remains a constant of integration to be determined from boundary conditions).

3. THE ENERGY-MOMENTUM TENSOR

Since minimal coupling is assumed for the final equilibrium state (i.e., the native particles and the monopoles interact only gravitationally), we may write a suitable energy-momentum tensor for each fluid species and combine these into one. For the native fluid, there is only the perfect-fluid tensor

$$(T_{\alpha\beta})_f = [\rho_f(r) + p_f(r)]u_\alpha u_\beta - p_f(r)g_{\alpha\beta} \qquad (3)$$

where, as usual, the (comoving) 4-velocity is defined as

$$u_0 = \sqrt{(g_{00})}, \quad u_\alpha u^\alpha = 1. \qquad (4)$$

For the monopole component, two source tensors are necessary. The first of these is just that of a perfect fluid,

$$(T_{\alpha\beta})_m = [\rho_m(r) + p_m(r)]u_\alpha u_\beta - p_m(r)g_{\alpha\beta}. \qquad (5)$$

Of course, the monostar's interior gravity must be strong enough to prevent the monopoles from diffusing through the native matter and evaporating out into space. Our model presupposes an equilibrium state such that the monopoles are distributed isotropically (though not necessarily homogeneously) throughout the monostar, their magnetic forces keeping them in suspension within the native fluid. We note in passing that setting $\rho_f(r) = p_f(r) = 0$ would describe a monostar composed of pure monopole fluid;

however, we are unaware of a physically plausible mechanism for the formation of an astrophysical object of this type and so will not pursue this possibility any further.

The second source term is that of the electromagnetic field,

$$(T_{\alpha\beta})_{\text{e-m}} = (1/4\pi)[-F_{\alpha\lambda}F_{\beta}^{\lambda} + \tfrac{1}{4}g_{\alpha\beta}F_{\lambda\nu}F^{\lambda\nu}], \tag{6}$$

built from the Maxwell tensor

$$F_{\alpha\beta} = A_{\beta,\alpha} - A_{\alpha,\beta} \tag{7}$$

where $A_\alpha$ is the 4-vector potential and a comma symbolizes an ordinary derivative. In the rest-frame of a spherical distribution of electric charge this potential reduces to $(A_\alpha)_\text{e} = [\varphi_\text{e}(r), 0, 0, 0]$, which gives the electric field strength as $E(r) \equiv -d\varphi_\text{e}/dr$.

To establish an analogous expression for a magnetic charge distribution, we turn to Kühne's symmetry postulate (Kühne, 1997): "The fundamental equations of the electromagnetic interaction describe electric and magnetic charges, electric and magnetic field strengths, and electric and magnetic potentials equivalently." This necessitates a second, magnetic, 4-vector potential of the same form as the electric one, which makes it possible to derive the symmetrized Maxwell equations proposed in the seminal paper on monopoles by Dirac (1931). By way of further justification, Kühne (1997) adds: "The second four-potential is required not only by the symmetry postulate, but also by the proven impossibility to construct a manifestly covariant one-potential model of quantum electromagnetodynamics."

Invoking the symmetry postulate, we define the magnetic 4-potential and magnetic field strength due to the monopoles as

$$(A_\alpha)_\text{m} \equiv [\varphi_\text{m}(r), 0, 0, 0], \quad B(r) = -d\varphi_\text{m}/dr. \tag{8}$$

From relation (7) the Maxwell tensor will now have the same form for magnetic monopoles as it would for electric charges, and so will the energy-momentum tensor (6) formed from it; it is therefore trivial to adapt the well-known electrostatic versions of $F_{\alpha\beta}$ and $(T_{\alpha\beta})_\text{e-m}$ to our magnetostatic model.

Finally, combining expressions (4), (5) and (6) and writing $\rho_\text{f} + \rho_\text{m} \equiv \rho$, $p_\text{f} + p_\text{m} \equiv p$ gives the total energy-momentum tensor for the interior of the monostar:

$$T_{\alpha\beta} = [\rho(r) + p(r)]u_\alpha u_\beta - p(r)g_{\alpha\beta} + (1/4\pi)[-F_{\alpha\lambda}F_{\beta}^{\lambda} + \tfrac{1}{4}g_{\alpha\beta}F_{\lambda\nu}F^{\lambda\nu}]. \tag{9}$$

4. THE EINSTEIN EQUATIONS

From the Einstein equations $G_\alpha^{\ \beta} = -8\pi T_\alpha^{\ \beta}$, we expect three independent relations for the four unknown functions $f(r)$, $\rho(r)$, $p(r)$ and $B(r)$. (For conciseness, we omit the $r$ dependence when writing these variables in equations from now on.) In addition, there is one nonzero relation from the Maxwell equations that introduces another unknown, the magnetic charge density $\lambda(r)$, which is expected to be functionally related to $B(r)$; even so, at some point an ansatz will have to be made for one of these unknown quantities.

Substituting relations (7) and (8) into the electromagnetic component of $T_{\alpha\beta}$, the Einstein equations reduce to:

$$(rf)' - 1 = -r^2(8\pi\rho + B^2) \tag{10}$$

$$(rf)' - 1 = r^2(8\pi p - B^2) \tag{11}$$

$$rf'' + 2f' = 2r(8\pi p + B^2) \tag{12}$$

where, as is usually done, a prime denotes $d/dr$. Subtracting relation (11) from (10) yields the Oppenheimer equation, which serves to give the relationship between pressure and density (the equation of state). A consequence of the initial choice of metric (1) is that $G_0{}^0 = G_1{}^1$ in the Einstein equations, making the left-hand sides of equations (10) and (11) identical; the Oppenheimer equation then becomes

$$p(r) = -\rho(r). \tag{13}$$

This shows that a perfect fluid (with or without electric or magnetic charge) may be described by a metric of the form (1) if and only if it has the equation of state defined by relation (13)—and, conversely, selecting this equation of state necessitates the specific form of metric given by expression (1).

If we require that $\rho > 0$, relation (13) states that the monostar's total internal pressure must be negative (making it a tension); specifically, $p = -\rho$ is the equation of state of a type of dark energy often called quintessence. (Recall that dark energy is defined by the inequality $-\rho \leq p \leq -\rho/3$, while $p < -\rho$ defines the phantom energy regime in which the Dominant Energy Condition, $\rho + p \geq 0$, no longer holds.) Physically, it is clear that a tension is necessary to counterbalance the magnetic field—which behaves like a positive pressure and, therefore, has an effective gravitational mass—in order to have equilibrium inside the monostar. This was established for an electric field by Tiwari, Kanakamedala & Rao (1984) as mentioned above, and prior to this by Cooperstock & de la Cruz (1978). We comment on possible physical origins of this tension below, in the Discussion.

An important relation that facilitates solution of the Einstein equations is the conservation equation, $T_\alpha{}^\beta{}_{;\beta} = 0$, where as usual the semicolon symbolizes a covariant derivative. Since our chosen metric yields only five independent, nonzero connection terms, the covariant divergence of the energy-momentum tensor (9) above reduces to

$$8\pi p' - (1/r^4)[r^4 B^2]' = 0 \tag{14}$$

after applying relation (13). This is in fact a generalization of the well-known Tolman-Oppenheimer-Volkoff (TOV) equation of stellar structure. Because expression (14) is obtainable from the Einstein equations (10) – (12) algebraically, it is not an additional independent equation; its value here lies in relating the pressure (and, thus, the density) to the magnetic field, allowing them to be found once $B(r)$ is known. It also emphasizes the important rôle of the magnetic field: Setting $B$ to zero would make the pressure function

unphysical—it would be constant throughout the monostar's interior, rather than the intuitively expected decreasing function of $r$ that vanishes at the surface.

5. THE INTERIOR SOLUTION

Of the two sets of Maxwell equations, one set, $F_{[\beta\gamma;\alpha]} = 0$, is automatically satisfied by the 4-potential (8) and the Maxwell tensor formed from it. The other set turns out to be the key to solving the entire system of Einstein-Maxwell equations describing the object's interior, because it leads to a simple and physically reasonable ansatz for the magnetic charge density $\lambda(r)$. This latter set of equations is $F^{\alpha\beta}{}_{;\beta} = 4\pi J^{\alpha}$, where $J^{\alpha} \equiv \lambda(r) u^{\alpha}$ is the magnetic 4-current density. The only nonzero component of these equations is

$$[r^2(-\varphi_m)']' = 4\pi r^2 \lambda(r) u^0 \qquad (15)$$

which, using relations (4) and (8) and integrating once, gives

$$r^2 B(r) = \int 4\pi r^2 [\lambda(r)/\sqrt{f(r)}] dr. \qquad (16)$$

The right-hand side becomes a straightforward volume integral of charge density (that is, the total magnetic charge $q_m$ contained within coordinate radius $r$) upon adopting the ansatz employed by Tiwari, Kanakamedala & Rao (1984) for the electrostatic case:

$$\lambda(r) = \lambda_0 \cdot \sqrt{f(r)}, \qquad (17)$$

where $\lambda_0$ is the (finite) charge density at the origin. Clearly, for this ansatz to be valid the metric function $f$ must also be finite at $r = 0$ and well-behaved all the way to the monostar's surface at coordinate radius $r = R$. Moreover, if $f(r)$ should turn out to be a decreasing function of $r$, it would validate the intuitive expectation that charge density should decrease with distance from the origin. (We see below that this is indeed the case.) Carrying out the integration in equation (16) and setting the constant of integration to zero yields

$$B(r) = (4/3)\pi r \lambda_0 \equiv q_m(r)/r^2, \qquad (18)$$

which establishes relation (17) as the simplest expression for a charge density giving an interior magnetic field that increases from the center to the surface. As expected, the exterior magnetic field will fall off as $1/r^2$ because the total magnetic charge contained within the monostar, $Q_m \equiv q_m(R)$, is a constant.

Inserting these expressions into the TOV equation (14) and integrating from the origin to the surface gives the pressure—and, from the Oppenheimer relation (13), the density—as

$$p(r) = -\rho(r) = (2/3)\pi \lambda_0^2 (r^2 - R^2). \qquad (19)$$

Note that pressure and density both vanish at the surface, and have finite values at the center of the object. To complete the interior solution, the Einstein equation (10) is

integrated once and the integration constant set to zero to avert a singularity at the origin, so that

$$f(r) = 1 - (1/r)\int r^2(8\pi\rho + B^2)dr. \tag{20}$$

Inserting appropriate expressions from equations (18) and (19) and carrying out the integral above finally gives the interior metric function as

$$f(r) = 1 - (16\pi^2/45)\lambda_0^2 r^2(5R^2 - 2r^2). \tag{21}$$

The interior metric will be well-behaved over the entire domain $0 \leq r \leq R$ as long as the function (21) remains positive; this requires that

$$R^2 < (15/16\pi^2)^{1/2}/\lambda_0. \tag{22}$$

The monostar's maximum (coordinate) radius is thus determined by its central magnetic charge density. Expressing the inequality (22) in terms of the central density $\rho_0$ via relation (19) and converting to SI units gives

$$R^2 < (5\pi c^2)/(8G\rho_0), \tag{23}$$

which allows us to estimate this radius for a given value of the central fluid density. For a typical neutron star of around two solar masses, this central density is believed to be around three times that of an atomic nucleus, or approximately $10^{18}$ kg m$^{-3}$. Taking this figure as the lower limit for the monostar's central density then gives $R < 5 \times 10^4$ m, which is at most of the same order of magnitude as the (true) radius of a neutron star. We certainly expect a greater central density for the monostar, but even increasing $\rho_0$ by two orders of magnitude in relation (23) still results in a plausible maximum coordinate radius of around $10^3$ m.

6. BOUNDARY CONDITIONS

Because the interior metric (1) and the exterior metric (2) have the same form, the boundary conditions at the monostar's surface reduce to requiring that the metric functions $f(r)$ and $\Phi(r)$ and their first derivatives satisfy

$$f(R) = \Phi(R); f'(r) \text{ and } \Phi'(r) \text{ continuous across } r = R. \tag{24}$$

The second of these conditions is easily established by direct calculation, and the first gives

$$(16\pi^2/15)\lambda_0^2 R^4 = r_S/R - Q_m^2/R^2. \tag{25}$$

Setting $r = R$ in relation (18) gives the monostar's global magnetic charge:

$$Q_m = (4/3)\pi R^3 \lambda_0. \tag{26}$$

Inserting this into expression (25) then yields the remaining unknown quantity of the model, the Schwarzschild radius:

$$r_S = (384/135)\pi^2 \lambda_0^2 R^5 = (8/5)Q_m^2/R. \qquad (27)$$

For the exterior metric to be well-behaved we must have $\Phi(r) > 0$, meaning that $r_S/r - Q_m^2/r^2 < 1$; the maximum value of the left-hand side of this expression will be at $r = R$, so using the result (27) it is easily seen that the inequality is satisfied for all $r \geq R$ as long as

$$(Q_m/R)^2 < 5/3. \qquad (28)$$

As expected, this is just a restatement of the condition (22) obtained by requiring the interior metric function $f(r)$ to be well-behaved. Assuming that the monostar stops collapsing before falling through its event horizon and attaining a quasi-black hole state requires that $R > r_S$. Again using relation (27), this requirement may be written as

$$(Q_m/R)^2 < 5/8, \qquad (29)$$

which is a more restrictive condition on the monostar's maximum radius than relation (28) or, equivalently, condition (22); it reduces the maximum value of $R$ to $(3/8)^{1/4} \approx 0.78$ of that obtainable from expressions (22) and (23).

Because the Schwarzschild radius is defined in terms of the object's total gravitational mass $M$ via the well-known relation $r_S = 2M$, from expression (27) we immediately have that

$$M = (4/5)Q_m^2/R. \qquad (30)$$

But $M = M_f + M_B$, where $M_f$ is the mass of the (native plus monopole) fluids comprising the monostar and $M_B$ is the effective gravitational mass of the monopolar magnetic field. The fluid mass is readily found from the mass function $\int 4\pi r^2 \rho(r) dr$ evaluated between the limits $r = 0$ and $r = R$; using relation (26), this gives $M_f = (1/5)Q_m^2/R$. We then obtain that $M_B = (3/5)Q_m^2/R$ and, consequently, that the ratio of the effective mass of the magnetic field to the mass of the fluid is

$$M_B/M_f = 3, \qquad (31)$$

which emphasizes the significance of the magnetic field's contribution to the monostar's gravitational properties.

Finally, relations (26) and (30) give the charge-to-mass ratio for the monostar:

$$Q_m/M = 15/(16\pi\lambda_0 R^2). \qquad (32)$$

Incorporating the condition (22) that the interior metric be well-behaved, modified by the more stringent requirement (29) that $R > r_S$, establishes a lower bound for the charge-to-mass ratio through the inequality

$$(Q_m/M)^2 > 5/2. \tag{33}$$

## 7. DISCUSSION

The phenomenon of negative internal pressure (tension) is an essential feature of several hypothetical compact-object and quasi-black hole models. As noted earlier, tension is a necessary property of any fluid source—charged or uncharged—described by a static, spherically symmetric metric where $g_{rr} = -1/g_{tt}$; it is required by the algebraic structure of the Einstein equations. For this geometric framework to be applicable to an object like the monostar means that, during the collapse process, a change of state must occur in the object's native fluid particles to produce the negative internal pressure. (We conjecture that mutual magnetic repulsion would serve to delay, or prevent, the onset of such a transition in the monopole population.) There is no known mechanism whereby such a change of state could take place at the nucleon stage, so this transition would presumably occur on a more fundamental level—if not in the quark domain, then among the hypothetical constituents of the quarks themselves, the preons.

Alternatively, a trivial modification of our energy-momentum tensor would allow us to invoke the transition of the quantum vacuum to a dark-energy-like state with $p = -\rho$ proposed by Mazur & Mottola (2004) as an essential feature of their model for a nonsingular final state of gravitational collapse, the gravitational vacuum condensate star. (The reader is referred to a review paper by Mottola (2010) for a thorough and detailed exposition of the "gravastar" hypothesis.) From this, we could speculate that if dark energy itself exists in the astrophysical regime, a rapidly rotating dark-energy star—which, of course, is under tension by definition—could become a form of monostar by capturing charged particles from the interstellar medium and thereby generating the dipolar magnetic field needed to produce magnetic monopoles as described in our Introduction.

Another issue is the empirical fact that no free monopoles have yet been detected, and it has been proposed that none may ever be found because they are too strongly confined in particle-antiparticle pairs named monopolium (Vento, 2008). If so, a collapsing compact object's dipolar magnetic field, however strong, may not be able to separate the monopole-antimonopole pairs produced by the proton-proton collisions; then, these particles would simply collect in the object's interior as monopolium. Devoid of the monopoles' signature magnetic field, the object would possess only the properties of a bi-fluid dark-energy star.

Finally, we consider the important issue of developing the monostar model in more detail, which would enable (among other things) the comparison of its physical properties with those of known compact objects. Inevitably, all theoretical models of as-yet unobserved entities reach the point where it is impossible to proceed further without empirically obtained values for fundamental physical quantities; in our case, however, just two such values would suffice to overcome this obstacle:

Recall that all of the functions describing the monostar—pressure, density, magnetic field, metric elements, and so on—explicitly depend on the object's central magnetic charge density $\lambda_0$. Since the monostar's exterior magnetic field is described by $B(r) = Q_m/r^2$, measuring the strength of this field at a known distance would immediately give the object's global magnetic charge $Q_m$. For a monostar located in a suitable binary system, we should also be able to determine its total gravitational mass $M$. Then, relation (30) would give the object's (coordinate) radius $R$, allowing the calculation of properties such as the moment of inertia and the Schwarzschild radius. Finally, inserting the value of $R$ into equation (32) would yield the value of $\lambda_0$, allowing the calculation of all other physical parameters needed to describe the monostar as an astrophysical entity.

## 8. CONCLUSION

Applying contemporary concepts from particle physics and the astrophysics of compact objects to a classical general relativistic framework, which was adapted and extended from one describing various hypothetical electrostatic entities, allowed us to describe the final equilibrium state of a stellar remnant infused with magnetic monopoles generated during its collapse; we proffered for this object the name, "monostar." Expressions for the object's global magnetic charge and Schwarzschild radius were obtained; then, the requirement that both interior and exterior metrics be free of singularities and horizons yielded inequalities delimiting the monostar's maximum radius and minimum magnetic charge-to-mass ratio.

The principal astrophysical signature of the monostar would, of course, be the radial, non-rotating, inverse-square magnetic field produced by the monopoles in its interior. We demonstrated that obtaining just two empirical results—the strength of this magnetic field at some exterior location, and the object's mass—would be all that is needed to derive values for the monostar's fundamental physical properties.

We conclude with the thought that, should it prove unfeasible to produce magnetic monopoles in particle colliders on Earth or to detect them as free particles in the interstellar medium, the discovery of an astrophysical object like the monostar would be a significant step towards establishing the existence and physical properties of these particles which were proposed by Dirac just over eighty years ago.